\documentclass{aa}
\usepackage[varg]{txfonts}
\usepackage{graphicx}
\usepackage{txfonts}
\usepackage{natbib}
\bibpunct{(}{)}{;}{a}{}{,} 
\begin{document}
\title{The    evolution    of     white    dwarfs    resulting    from
  helium-enhanced, low-metallicity progenitor stars.}

\author{Leandro G. Althaus\inst{1,2}, 
        Francisco C. De Ger\'onimo\inst{1,2}, 
        Alejandro H. C\'orsico\inst{1,2}, 
        Santiago Torres\inst{3,4}, \and  
        Enrique Garc\'\i a--Berro\inst{3,4}}
\institute{Grupo de Evoluci\'on Estelar y Pulsaciones. 
           Facultad de Ciencias Astron\'omicas y Geof\'{\i}sicas, 
           Universidad Nacional de La Plata, 
           Paseo del Bosque s/n, 1900 
           La Plata, 
           Argentina
           \and
           IALP - CONICET
           \and
           Departament de F\'\i sica, 
           Universitat Polit\`ecnica de Catalunya, 
           c/Esteve Terrades 5, 
           08860 Castelldefels, 
           Spain
           \and
           Institute for Space Studies of Catalonia,  
           c/Gran Capita 2--4, Edif. Nexus 201, 
           08034 Barcelona,
           Spain}
\date{Received ; accepted }

\abstract{Some  globular clusters  host  multiple stellar  populations
           with  different  chemical   abundance  patterns.   This  is
           particularly true for $\omega$  Centauri, which shows clear
           evidence of a  helium-enriched sub-population characterized
           by a helium abundance as high as $Y= 0.4$}
         {We present a whole and consistent set of evolutionary tracks
           from the ZAMS to the  white dwarf stage appropriate for the
           study  of  the  formation  and evolution  of  white  dwarfs
           resulting from the evolution of helium-rich progenitors.}
         {White  dwarf sequences  have been  derived from  progenitors
           with stellar mass ranging from 0.60 to $2.0\, M_{\sun}$ and
           for an initial helium abundance  of $Y= 0.4$.  Two values of
           metallicity have been adopted: $Z= 0.001$ and $Z= 0.0005$.}
         {Different  issues of  the  white dwarf  evolution and  their
           helium-rich progenitors have  been explored. In particular,
           the final  mass of the  remnants, the role  of overshooting
           during the thermally-pulsing phase,  and the cooling of the
           resulting   white   dwarfs   differ   markedly   from   the
           evolutionary predictions of  progenitor stars with standard
           initial   helium  abundance.    Finally,  the   pulsational
           properties  of   the  resulting   white  dwarfs   are  also
           explored.}
         {We find  that, for the  range of initial masses  explored in
           this paper,  the final mass of  the helium-rich progenitors
           is  markedly  larger  than  the final  mass  expected  from
           progenitors with the usual  helium abundance.  We also find
           that progenitors  with initial  mass smaller  than 
           $M_{\star}\simeq
           0.65\,M_{\sun}$  evolve  directly  into  helium-core  white
           dwarfs in less than 14~Gyr,  and that for larger progenitor
           masses   the   evolution    of   the   resulting   low-mass
           carbon-oxygen white dwarfs is dominated by residual nuclear
           burning.  For  helium-core white dwarfs, we  find that they
           evolve markedly faster than  their counterparts coming from
           standard progenitors.   Also, in contrast with  what occurs
           for  white  dwarfs  resulting  from  progenitors  with  the
           standard helium  abundance, the impact of  residual burning
           on the cooling time of white  dwarfs is not affected by the
           occurrence  of  overshooting during  the  thermally-pulsing
           phase of progenitor stars.}
\titlerunning{White  dwarf  evolutionary   sequences  for  He-enhanced
 progenitors}
\authorrunning{Althaus et al.}

\maketitle


\section{Introduction}
\label{introduction}

It is now well established that a handful of globular clusters harbour
multiple   stellar   populations   characterized   by   different   He
enrichments \citep{2015MNRAS.449.3333B}. This is the case for
instance for $\omega$~Centauri   \citep{2004ApJ...612L..25N,
  2013ApJ...762...36J, 2016MNRAS.457.4525T},     NGC     2808
\citep{2005ApJ...631..868D, 2012ApJ...754L..34M},   and   NCG   6441
\citep{2007A&A...463..949C, 2013ApJ...765...32B}  in which
sub-populations  are clearly  visible  in their respective
color-magnitude diagrams.  In particular, He enrichments as high as
$Y\sim 0.4$ have been  suggested  to explain  the  observed  split
main sequences  and Horizontal  Branches   (HB)  of $\omega$~Centauri
and NGC     2808.  Because an increase in He abundance  decreases the
turnoff mass  of a  stellar population,  the presence  of a
He-enriched sub-population  has also  been invoked  to account   for
the   presence   of   hot   and   extreme   HB   stars
\citep{2008MNRAS.390..693D},  as  well as  the  existence  of a  large
fraction  of He-core  white  dwarfs in  some  globular clusters,  like
$\omega$~Centauri  \citep{2008ApJ...673L..29C, 2013ApJ...769L..32B}.
The  origin of  these He-rich   sub-populations  is   believed   to
result  from   cluster self-pollution  caused  by  the ejecta  of
massive-intermediate  mass Asymptotic Giant Branch (AGB) stars  of the
first generation after the occurrence  of  the second  dredge-up  and
hot bottom  burning  stage \citep{2001ApJ...550L..65V},   from   fast
rotating   massive   stars \citep{2007A&A...464.1029D}, or  from
evolved  Red Giant  Branch (RGB) stars that experienced extra-deep
mixing \citep{2004ApJ...603..119D} ---  see also
  \cite{2015MNRAS.449.3333B} for a recent discussion about this
  issue.

The detection of white dwarf cooling sequences of globular clusters is
specially interesting, since it allows studying fundamental properties
of these stars and of  the corresponding stellar populations.  Because
of     their      well     understood      evolutionary     properties
\citep{2008PASP..120.1043F,2008ARA&A..46..157W,  2010A&ARv..18..471A},
white dwarfs  can be  used as distance  indicators and  as independent
reliable cosmic clocks to date  a wide variety of stellar populations,
such as our  Galaxy -- see \cite{wdlf}, and references  therein, for a
recent review  on this topic --  and open and globular  clusters --
see \cite{2009ApJ...693L...6W},
\cite{2010Natur.465..194G}, \cite{2011ApJ...730...35J},
\cite{2013A&A...549A.102B}       and \cite{2013Natur.500...51H} for
some examples.

Recently,  \cite{2015A&A...576A...9A} have  explored the  evolution of
white dwarfs formed  in metal-poor populations, and  found that stable
hydrogen  burning dominates  their cooling  even at  low luminosities.
They  found  that  the  role   played  by  such  residual  burning  is
independent of the  rate at which stellar mass is  lost during the AGB
and post-AGB evolution  of progenitors, but depends  on the occurrence
of  overshooting  during the  thermally-pulsing  (TP)  AGB phase.   In
particular,  overshooting  results  in  a  carbon  enrichment  of  the
envelope  due   to  third   dredge-up  episodes.    Carbon  enrichment
eventually  results in  final  thinner hydrogen  envelopes, which  are
unable to sustain appreciable hydrogen  burning during the white dwarf
stage.    Here,    we   extend   the    scope   of   the    study   of
\cite{2015A&A...576A...9A}  to  explore  the   impact  of  an  initial
He-enrichment on  the formation, evolution and  pulsational properties
of the resulting white dwarf stars.

Our main  aim is to  provide a  consistent set of  evolutionary tracks
from the ZAMS to the cooling  phase appropriate for the study of white
dwarfs  with   helium  and  carbon-oxygen  cores   formed  in  He-rich
sub-populations. The  white dwarf evolutionary sequences  are computed
from the  full evolution of  He-rich progenitor stars through  all the
relevant stellar evolutionary phases, including the ZAMS, the RGB, the
core He  flash (whenever it occurs),  the stable core He  burning, the
AGB phase  and the entire  TP- and post-AGB phases.   Specifically, we
present  full  evolutionary  sequences for  He-rich  progenitors  with
masses ranging  from 0.6 to  $2.0\, M_{\sun}$,  and for an  initial He
abundance $Y$=0.4.   These sequences  are provided for  two progenitor
metallicities $Z$=0.001  and $Z$=0.0005.  To date,  no such
exploration of the evolution  of He-rich stars exists in the
literature, the only exception  is   that  of
\cite{2015A&A...578A.117C},   who  presented evolutionary sequences
for He-rich  progenitors for low-mass stars (up to $1\, M_{\sun}$)
with $Z$=0.0005 from the ZAMS to the end of the AGB phase, but did
not compute the white dwarf stage.   We emphasize that the
comptutation of  the  entire evolutionary  history of  progenitor
stars allows  us to have  self-consistent white dwarf  initial models.
That means  that in our  sequences the  residual masses of  the H-rich
envelopes  and  of  the  He  shells  are  obtained  from  evolutionary
calculations, instead  of using typical values  and artificial initial
white dwarf models.  For the white  dwarf regime, we have included all
the   relevant  energy   sources  and   physical  processes   such  as
crystallization,  carbon-oxygen phase  separation, element  diffusion,
residual nuclear burning, and  convective coupling at low luminosities
\citep{2001PASP..113..409F}.   As it  will be  shown below,  the final
mass  of   the  remnants,   the  role   of  overshooting   during  the
thermally-pulsing  AGB, and  the  cooling properties  of white  dwarfs
markedly differ from those corresponding  to progenitor stars with the
standard initial He abundance.

The paper  is organized  as follows.   In Sect.~\ref{code}  we briefly
describe  our  numerical  tools  and   the  main  ingredients  of  the
evolutionary  sequences, while  in Sect.~\ref{results}  we present  in
detail our  main evolutionary  results for both  the white  dwarfs and
their progenitor.   In this  section we  also explore  the pulsational
properties  of   the  resulting   white  dwarfs  during   the  ZZ~Ceti
stage.  Finally,  in  Sect.~\ref{conclusions} we  summarize  the  main
findings of the paper, and we elaborate on our conclusions.

\begin{table*}
\caption{Basic  model  properties  for   sequences  with  $Y$=0.4  and
  $Z$=0.001 and 0.0005.}
\centering
\begin{tabular}{lcclc}
\hline
\hline
$M_{\rm ZAMS}\, (M_{\sun})$ & $t_{\rm RGB}$~(Gyr) & $M_{\rm WD}\, (M_{\sun})$ &
 $N_{\rm TP}$ & Evolutionary path\\
\hline
\multicolumn{4}{c}{$Y$=0.4, $Z$=0.001}\\
\hline
0.60  &  14.15 & 0.4336 & 0 & He-core\\ 
0.65  & 10.57 & 0.450 & 0 &He-core\\ 
0.70  & 8.08 & 0.4894 & 4 (at high $T_{\rm eff}$) &  Hot HB $\rightarrow$ AGB Manqu\`e $\rightarrow$ C/O WD  \\ 
0.75  & 6.31 & 0.52801 & 2 (at high $T_{\rm eff}$) & HB $\rightarrow$ TP-AGB Manqu\`e $\rightarrow$ C/O WD  \\         
0.80  &  5.02 & 0.55837 & 2 (1 at high $T_{\rm eff}$) & HB  $\rightarrow$  TP-AGB   $\rightarrow$ C/O WD  \\          
0.85  & 4.09 & 0.57982 & 2 & HB  $\rightarrow$  TP-AGB  $\rightarrow$ C/O WD  \\  
1.0   & 2.35 & 0.61561 & 4 & HB  $\rightarrow$   TP-AGB $\rightarrow$ C/O WD  \\ 
1.5   & 0.648 & 0.70424 & 11 &  TP-AGB $\rightarrow$ C/O WD  \\ 
2.0   &  0.31 & 0.81114 & 20 & TP-AGB  $\rightarrow$ C/O WD \\
\hline
\multicolumn{4}{c}{$Y$=0.4, $Z$=0.0005}\\
\hline
0.60  &  13.49 & 0.43968 & 0 &  He-core\\ 
0.65  & 10.08 & 0.45746 & 0 & He-core\\ 
0.70  & 7.72 & 0.50053 &  4 (at high $T_{\rm eff}$) & Hot HB $\rightarrow$ TP-AGB Manqu\`e $\rightarrow$ C/O WD  \\ 
0.75  & 6.038 & 0.53385 & 3 (1 at high $T_{\rm eff}$) & HB $\rightarrow$ TP-AGB  $\rightarrow$ C/O WD  \\         
0.85  & 3.893 & 0.58029 & 3 &  HB $\rightarrow$ TP-AGB $\rightarrow$ C/O WD  \\  
1.0   & 2.235 & 0.62152 & 5 & HB $\rightarrow$ TP-AGB $\rightarrow$ C/O WD  \\ 
1.5   & 0.626 & 0.71232 & 12 &   TP-AGB $\rightarrow$ C/O WD  \\ 
2.0   & 0.293 & 0.83789 & 24 &  TP-AGB $\rightarrow$ C/O WD \\ 
\hline
\end{tabular}
\tablefoot{$M_{\rm ZAMS}$: initial mass, $t_{\rm RGB}$: age at the RGB
  (in Gyr),  $M_{\rm WD}$: white  dwarf mass, $N_{\rm TP}$:  number of
  thermal pulses. The last column gives the evolutionary path followed
  by the star.}
\label{tabla1}
\end{table*}

\section{Numerical setup and input physics}
\label{code}

The evolutionary  sequences presented  here have been  calculated with
the    {\tt    LPCODE}    stellar    evolutionary    code    --    see
\cite{2003A&A...404..593A},                \cite{2005A&A...435..631A},
\cite{2012A&A...537A..33A},       \cite{2015A&A...576A...9A},      and
\cite{2016A&A...588A..25M}  for relevant  information about  the code.
{\tt LPCODE} is a well-tested and  calibrated code that has been amply
used to study different aspects of the evolution of low-mass and white
dwarf  stars,  including  the  formation and  evolution  of  extremely
low-mass    white    dwarfs   --    see    \cite{2008A&A...491..253M},
\cite{2010Natur.465..194G},                \cite{2010ApJ...717..897A},
\cite{2010ApJ...717..183R},                \cite{2011ApJ...743L..33M},
\cite{2011A&A...533A.139W},       \cite{2012MNRAS.424.2792C},      and
\cite{2013A&A...557A..19A}, and references  therein.  Recently, it has
been used as well to generate a  new grid of models for post-AGB stars
\citep{2016A&A...588A..25M}.  {\tt  LPCODE}  has been  tested  against
other  evolutionary codes  during the  main sequence,  RGB, and  white
dwarf   regime  \citep{2013A&A...555A..96S,2016A&A...588A..25M}   with
satisfactory results.

Next we  provide a description of  the main input physics  of the code
relevant  for  the  present  work.   Extra  mixing  due  to  diffusive
convective overshooting has  been considered during the core  H and He
burning, but not during the  thermally-pulsing AGB phase.  The nuclear
network  accounts  for  the   following  elements:  $^{1}$H,  $^{2}$H,
$^{3}$He, $^{4}$He, $^{7}$Li,  $^{7}$Be, $^{12}$C, $^{13}$C, $^{14}$N,
$^{15}$N,  $^{16}$O,  $^{17}$O,   $^{18}$O,  $^{19}$F,  $^{20}$Ne  and
$^{22}$Ne,  together  with 34  thermonuclear  reaction  rates for  the
pp-chains, CNO  bi-cycle, and He  burning that are identical  to those
described  in \cite{2005A&A...435..631A},  with the  exception of  the
reactions  $^{12}$C$\   +\  $p$   \rightarrow  \  ^{13}$N   +  $\gamma
\rightarrow    \    ^{13}$C    +    e$^+   +    \nu_{\rm    e}$    and
$^{13}$C(p,$\gamma)^{14}$N,       which      are       taken      from
\cite{1999NuPhA.656....3A}.    In    addition,   the    reacion   rate
$^{14}$N(p,$\gamma)^{15}$O was  taken from \cite{2005EPJA...25..455I}.
Radiative and conductive opacities are  computed usint the OPAL tables
\citep{1996ApJ...464..943I}    and   adopting    the   treatment    of
\cite{2007ApJ...661.1094C},  respectively.   The   equation  of  state
during the main sequence evolution is  that of OPAL for H- and He-rich
compositions  and   a  given  metallicity.    Updated  low-temperature
molecular opacities with varying  carbon-oxygen ratios are used, which
is relevant for  a realistic treatment of  progenitor evolution during
the thermally-pulsing AGB phase \citep{2009A&A...508.1343W}.  For this
purpose,   we   have   adopted  the   low-temperature   opacities   of
\cite{2005ApJ...623..585F}  and  \cite{2009A&A...508.1343W}.  For  the
white   dwarf  phase,   we   consider  the   equation   of  state   of
\cite{1979A&A....72..134M} for  the low-density regime, while  for the
high-density   regime,   we   employ   the  equation   of   state   of
\cite{1994ApJ...434..641S}. As the white dwarf cools down we take into
account  the  effects  of   element  diffusion  due  to  gravitational
settling,  chemical and  thermal diffusion  of $^1$H,  $^3$He, $^4$He,

$^{12}$C,     $^{13}$C,    $^{14}$N     and     $^{16}$O    --     see
\cite{2003A&A...404..593A} for details.  During the white dwarf regime
and for effective  temperatures smaller than 10,000  K, outer boundary
conditions    are   derived    from    non-grey   model    atmospheres
\citep{2012A&A...546A.119R}.   Energy sources  resulting from  nuclear
burning, the release of latent heat of crystallization, as well as the
release  of the  gravitational energy  associated with  the C/O  phase
separation  induced by  crystallization  are also  taken into  account
during this phase.

In the present work special care  has been considered in computing the
evolution  along the  TP-AGB phase.  This  is relevant  for a  correct
assessment  of the  initial-to-final mass  relation as  well as  for a
realistic inference  of the role  of residual nuclear burning  in cool
white dwarfs \citep{2015A&A...576A...9A}.  In  particular, we have not
forced our sequences to abandon early the TP-AGB phase.  As mentioned,
overshooting was not taken into account during the TP-AGB phase.  This
assumption leads to  upper limits to the final mass  of the progenitor
stars.   Indeed,  the  inclusion  of overshooting  during  this  phase
results in  third dredge up episodes,  that prevent the growth  of the
H-free  core \citep{2009ApJ...692.1013S}.   Unlike  the situation  for
metal-rich progenitors,  evidence for  the occurrence  of extra-mixing
episodes during the TP-AGB phase of low-mass, low-metallicity stars is
not conclusive  \citep{2015A&A...576A...9A}. To  assess the  impact of
the occurrrence of  extra-mixing in the TP-AGB  phase, particularly on
the final mass of progenitors and the role of residual nuclear burning
during the white dwarf regime, we compute additional He-rich sequences
in which we considered an exponentially decaying diffusive overhooting
with  the  overshooting parameter  set  to  $f$=0.0075.  As  shown  in
\cite{2016A&A...588A..25M},  this  amount of  overshooting  reproduces
many observational  properties of AGB  and post-AGB stars such  as the
C/O  ratios of  AGB  and  post-AGB stars  in  the  Galactic Disk,  C/O
abundances in PG~1159 stars, and the mass range of C-rich stars in the
clusters of the Magellanic Clouds.

In   this  work,   mass   loss   during  the   RGB   was  taken   from
\cite{2005ApJ...630L..73S}. Although this election is an acceptable choice, it should be
  stressed, however, that the RGB mass-loss law is uncertain, and 
  particularly its  dependence on the initial
helium abundance is still unknown. For the  AGB and TP-AGB phases, we use
again the  mass loss rate of  \cite{2005ApJ...630L..73S} for pulsation
periods shorter than  50~days. For longer periods, mass  loss is taken
as  the  maximum  of   the  rates  of  \cite{2005ApJ...630L..73S}  and
\cite{2009A&A...506.1277G} for  oxygen-rich stars,  or the  maximum of
the rates of \cite{2005ApJ...630L..73S} and \cite{1998MNRAS.293...18G}
for  carbon-rich stars.   In all  of our  calculations, mass  loss was
suppressed after the post-AGB remnants reach $\log T_{\rm eff}$=4.  In
line  with the  findings of  \cite{2014A&A...571A..81C}, who  reported
that no enhanced mass loss should  be expected in second generation HB
stars, we assume no mass loss  during the core He-burning phase on the
HB.

We  computed the  full  evolution of  He-rich  sequences with  initial
masses ranging from 0.6 to $2.0\, M_{\sun}$.  The initial He abundance
of each  sequence is $Y$=0.4.  Two  sets of model sequences  have been
computed,  one  for  metallcity  $Z$=0.001   and  the  other  one  for
$Z= 0.0005$.  It is important  to mention  that the value   of the
initial helium abundance adopted here constitutes an extreme
choice. In fact,  only $\omega$ Centauri and NGC~2808 show evidence of
sub-populations  with such high large helium abundances,   while  the
sub-populations  of most typical clusters  are characterized by helium
abundances  generally  $\la 0.3$ \citep{2015MNRAS.449.3333B}. A
description of the  computed  evolutionary sequences is provided in
Table~\ref{tabla1}, which lists  the initial mass  of our sequences
and the  final white dwarf mass (in  solar masses), together with the
age (in Gyr) at the end  of the RGB and the number of thermal pulses.
The adopted metallicities are representative of the low-metal content
of  the second  generation, He-rich  population found  in some
globular  clusters.   In particular,  \cite{2016MNRAS.457.4525T}  find
that the metallicity of the He-rich population in $\omega$ Centauri is
$Z= 0.0006-0.001$, and  $Y\sim 0.38$.  For all our  sequences, evolution
has  been computed  starting from  the ZAMS  and followed  through the
stages of stable H and He core  burning, the stage of mass loss during
the  entire  TP-AGB, the  domain  of  the  planetary nebulae  at  high
effective temperature,  and finally  the terminal white  dwarf cooling
track, until very  low surface luminosities ($\log(L/L_{\sun})= -5.0$).
For most  of our  sequences, evolution  has gone  through the  He core
flash on the tip of the  RGB, and the following recurrent sub-flashes,
before the progenitors  reach the stable core He-burning  stage on the
HB.

\section{Evolutionary results}

\label{results}

In this work we will focus on  those aspects of the evolution of stars
with enhanced  initial helium  abundances which  are relevant  for the
formation and  evolution of white  dwarf stars.  Other issues  such as
the impact  of the helium enhancement  of the progenitor stars  on the
morphology of  HB stars, on  the existence of multiple  populations in
globular clusters, or on the consequences of the helium enhancement on
the  evolution and  nucleosynthesis  of  AGB models  can  be found  in
\cite{2002A&A...395...69D},                \cite{2013A&A...557L..17C},
\cite{2015MNRAS.446.1672M}, \cite{2015MNRAS.452.2804S}, and references
therein.

\begin{figure}
\centering
\includegraphics[clip,width=\columnwidth]{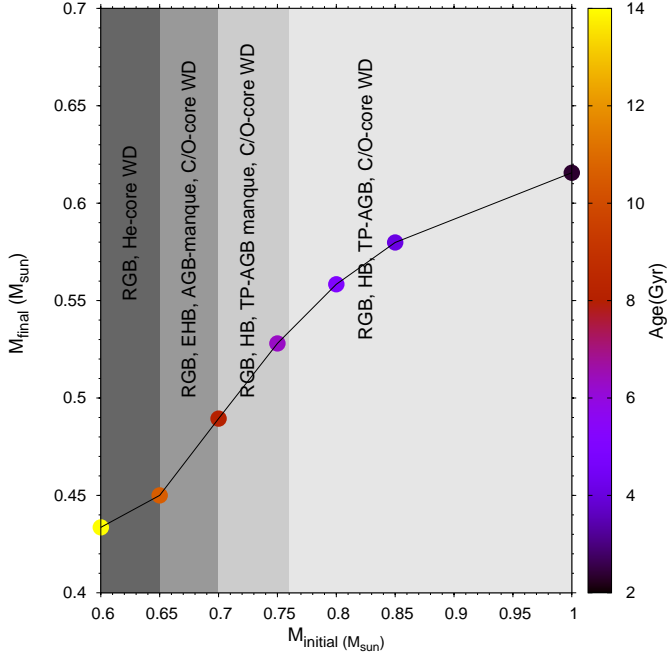}
\caption{Theoretical    initial-to-final    mass   relationship    and
  evolutionary stages (shown employing different gray scales) obtained
  following  the evolution  of stars  with an  initial helium-enhanced
  composition $Y$=0.4 and a metallicity $Z$=0.001.  The color scale on
  the right shows the age (in Gyr) at the tip of the RGB.}
\label{mimfY04Z1d3}
\end{figure}

\subsection{The evolution of white dwarf progenitors}

It is  well known that  stars with  an enhanced helium  abundance have
much shorter  main sequence  lifetimes for a  given initial  mass than
stars  with  a   normal  helium  abundance.   This   is  the  logical
consequence  of  a  smaller  hydrogen  mass  to  burned  on  the  main
sequence.   However   this  is   not   the   only  reason   for   this
behavior.  Specifically,  in  the   hydrogen  burning  core  the  mean
molecular weight  is larger.  All in all,  stars with  enhanced helium
abundances   are  brighter   and  hotter   for  a   given  mass,   see
\cite{2005essp.book.....S} and  \cite{2013A&A...557L..17C}. Indeed, as
can be  noted in Table~\ref{tabla1}, progenitors  with initial stellar
masses as low as $M\simeq 0.60\,M_{\sun}$  reach the tip of the RGB in
less  than 14~Gyr.   In addition,  because of  the larger  temperature
characterizing the hydrogen-exhausted core  during the RGB, the helium
flash for  these stars  occurs for lower  helium core masses  than for
stars with normal helium abundances.

These   facts  have   consequences  for   the  initial-to-final   mass
relationship,  as well  as for  the global  evolution of  these stars.
This   is    illustrated   in   Fig.~\ref{mimfY04Z1d3},    where   the
initial-to-final mass  relation for  our less  massive helium-enhanced
evolutionary  sequences is  shown  for the  case  $Z$=0.001.  In  this
figure the evolution  paths of these stars according  to their initial
mass is also  indicated. In particular, progenitors  with initial mass
less  than $M\simeq0.65\,M_{\sun}$  do not  evolve through  the helium
core flash at the tip of the RGB. Instead, they evolve directly to the
white dwarf  stage.  For  these progenitors, helium-core  white dwarfs
are  formed  in  less  than  14~Gyr  (see  Table  \ref{tabla1}).   Our
sequences   with  initial   stellar   masses   larger  than   $M\simeq
0.65\,M_{\sun}$ end their  lives forming white dwarfs  with C/O cores.
Specifically, helium-enhanced stars with initial stellar masses within
the range  $0.65\la M_{\sun}\la  1.0$ evolve  through the  helium core
flash and then  to the HB.  It is worth  noting that those progenitors
with masses in the interval $0.65 \la M_{\sun} \la 0.70$ avoid the AGB
phase (AGB manqu\`e),  and evolve directly to the  cooling sequence as
C/O  white  dwarfs --  see  \cite{2015A&A...578A.117C}  for a  similar
result --  while those progenitors  with $0.70 \la M_{\sun}  \la 0.75$
reach the  AGB phase but not  the thermally pulsing AGB  phase (TP-AGB
manqu\`e).

\begin{figure}
\centering
\includegraphics[clip,width=\columnwidth]{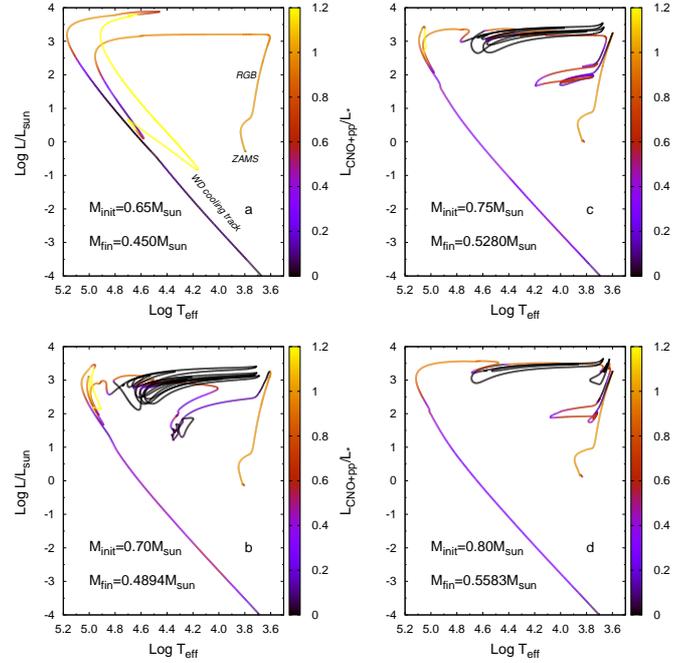}
\caption{Hertzsprung-Russell    diagram   for    our   helium-enhanced
  evolutionary sequences for metallicity  $Z$=0.001 and initial helium
  mass  fraction $Y=0.4$.   Evolutionary tracks  for progenitors  with
  initial  stellar  mass  of  $0.65\,  M_{\sun}$,  $0.70\,  M_{\sun}$,
  $0.75\,  M_{\sun}$,  and $0.80\,  M_{\sun}$  (panels  a,b,c, and  d,
  respectively) are depicted from the ZAMS to advanced stages of white
  dwarf evolution. The color scale on  the right shows the fraction of
  the total luminosity due to hydrogn burning.  The upper value of the
  color scale has been  set to 1.2, so larger values  of $L_{\rm CNO +
  pp}/L_*$ are not shown.}
\label{hr_Y04Z1d3}
\end{figure}

\begin{figure}
\centering
\includegraphics[clip,width=\columnwidth]{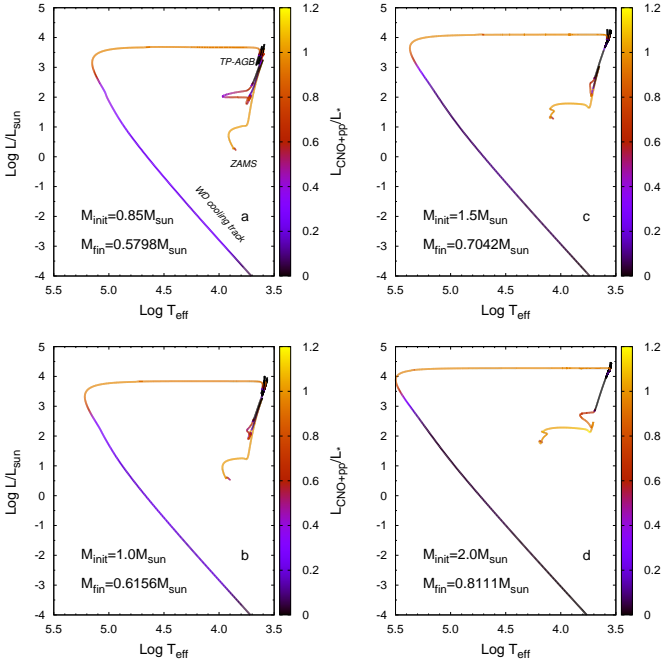}
\caption{Same  as  Fig.~\ref{hr_Y04Z1d3}   but  for  the  evolutionary
  sequences of stars with  masses $0.85\, M_{\sun}$, $1.0\, M_{\sun}$,
  $1.5\, M_{\sun}$, and $2.0\, M_{\sun}$. }
\label{hr_Y04Z1d3_masivo}
\end{figure}

In our simulations,  only those progenitors with  initial stellar mass
larger than $\approx 0.70\, M_{\sun}$ are able to reach the AGB phase.
This    is   in    good   agreement    with   the    calculations   of
\cite{2013A&A...557L..17C}.    This   is    clearly   illustrated   in
Figs.~\ref{hr_Y04Z1d3} and \ref{hr_Y04Z1d3_masivo},  which display the
evolution  in   the  Hertzsprung-Russell  diagram  of   some  selected
sequences with $Z$=0.001, from the ZAMS  to the white dwarf stage. The
color scale in each figure shows  the fraction of the total luminosity
due  to hydrogen  burning. Note  that the  sequence with  initial mass
$M=0.70\,M_{\sun}$ (panel b in Fig. \ref{hr_Y04Z1d3}) evolves directly
to the white  dwarf regime without passing through the  AGB phase (AGB
manqu\`e).  Indeed,  this sequence,  that experiences the  core helium
flash shortly after departing from the RGB phase, settles onto the hot
HB (at $T_{\rm eff}=22,400K$) to burn helium in a stable way.  Because of
its very thin hydrogen envelope, this sequence does not go through the
AGB phase  after the end  of core helium  burning.   On  the other  hand, the sequence
with initial
mass $M=0.75\,M_{\sun}$ (panel c in Fig.~\ref{hr_Y04Z1d3}) reaches the
AGB  but  abandons  it  before   reaching  the  TP-AGB  phase  (TP-AGB
manqu\`e).   However,  note  that both  sequences  experience  several
helium shell flashes before reaching the white dwarf regime.

Panel a in  Fig.~\ref{hr_Y04Z1d3} shows the complete  evolution of our
progenitor star  of mass  $M=0.65\,M_{\sun}$ that ends  its life  as a
helium-core white  dwarf of mass $M=0.45\,M_{\sun}$.   Before reaching
the  terminal cooling  track, this  sequence experiences  a CNO  shell
flash, that  reduces the effects  of residual hydrogen burning  on its
evolution  at  late cooling  times.   This  is  in contrast  with  the
expectation for  very massive helium-core white  dwarfs resulting from
progenitors with  standard initial  helium abundances.  In  this case,
CNO   flashes   are   not   expected  to   occur,   as   reported   in
\cite{2002MNRAS.337.1091S}    and   \cite{2013A&A...557A..19A}.    The
occurrence or  not of a  CNO shell flash  is critical for  the cooling
times of  these stars.  Finally, note  from Figs.~\ref{hr_Y04Z1d3} and
\ref{hr_Y04Z1d3_masivo} that, except for  the more massive white dwarf
sequences  and for  the  helium-core ones,  residual hydrogen  burning
constitutes a main  energy source for all the  resulting white dwarfs,
also at advanced stages of evolution (see later in this section).

\begin{figure}
\centering
\includegraphics[clip,width=\columnwidth]{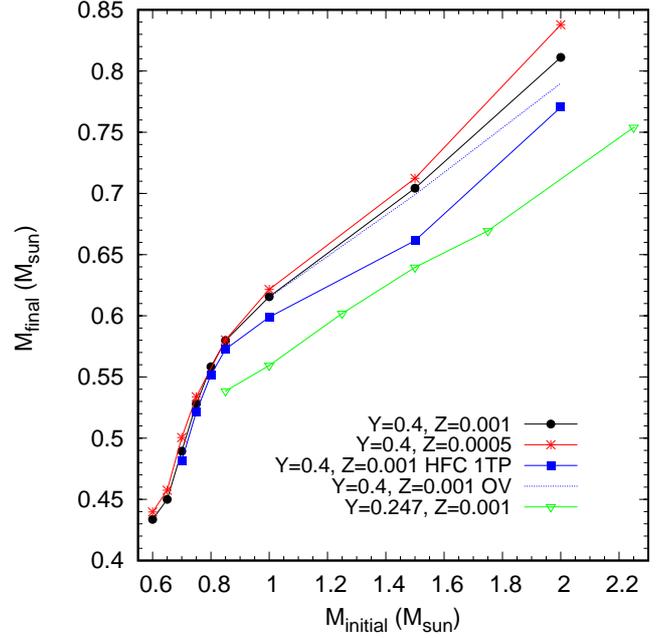}
\caption{Theoretical   initial-to-final   mass    relation   for   our
  helium-enhanced  evolutionary sequences  with metallicity  $Z$=0.001
  and $Z$=0.0005  (black and red  solid lines, respectively).  We also
  show the  mass of the  HFC at  the first thermal  pulse for  the case
  $Y$=0.4,   $Z$=0.001   (blue   solid   line).   In   addition,   the
  initial-to-final  mass  relation  when  overshooting  is  considered
  during  the TP-AGB  phase, and  that resulting  from sequences  with
  standard  helium content  ($Y$=0.247 and  $Z$=0.001) are  also shown
  (blue dotted and green solid lines, respectively).}
\label{mimfull}
\end{figure}

The impact of  the initial helium content on the  resulting final mass
is  shown  in  Fig.~\ref{mimfull},   which  displays  the  theoretical
initial-to-final  mass  relation  resulting  for  our  helium-enhanced
evolutionary sequences  for metallicity $Z$=0.001 and  $Z$=0.0005. The
final mass is the stellar mass with which the remnant enters the white
dwarf phase. The figure also shows  the mass of the hydrogen-free core
(HFC) at  the first thermal  pulse for the  case in which  $Y$=0.4 and
$Z$=0.001 are adopted.  For the sake of comparison, we also include in
the figure the  initial-final-mass relation for sequences  that do not
consider a helium enhancement.  In  particular, we plot, using a green
solid  line,  the  resulting  relationship  for  the  case  $Y$=0.247,
$Z$=0.001  \citep{2015A&A...576A...9A}.  The  growth of  the HFC  mass
during the  TP-AGB, i.e.  beyond  the first thermal pulse,  is evident
for initial  masses larger than $M\sim1.0\,M_{\sun}$.   Because of the
small amount  of mass  remaining above  the HFC  at the  first thermal
pulse, the final  mass for the less massive sequences  does not differ
appreciably from the mass of the HFC at the first thermal pulse.

\begin{figure}
\centering
\includegraphics[clip,width=\columnwidth]{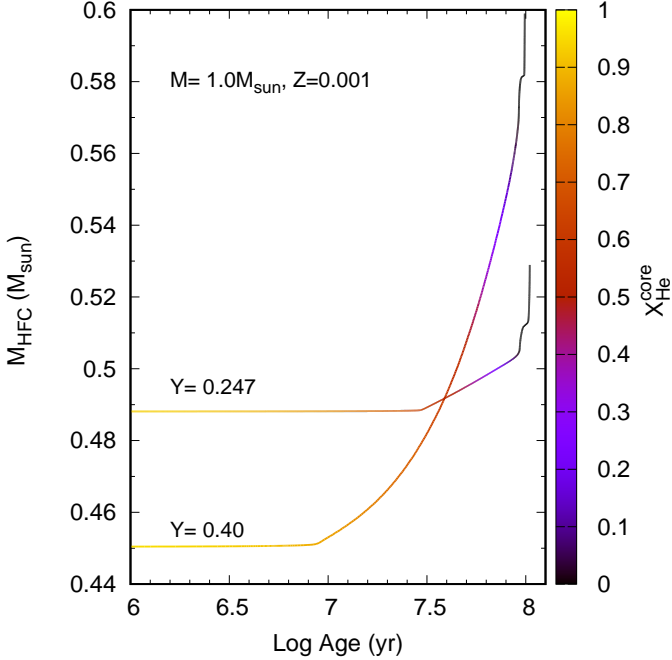}
\caption{Time evolution of the mass of  the HFC from the onset of core
  helium burning until  the occurrence of the first  thermal pulse for
  the  $  1.0  \,M_{\sun}$,   $Z$=0.001  sequences  with  $Y$=0.4  and
  $Y$=0.247.  The color  scale  on  the right  shows  the core  helium
  abundance (mass fraction). Note the  substantial increase of the HFC
  mass during core helium burning for the helium-enhanced sequence.}
\label{hfc}
\end{figure}

It  is  also  worth  noting  in  this  figure  that  in  the  case  of
helium-enhanced sequences, the final mass  of the remnants is markedly
larger than  the final  mass expected from  sequences with  a standard
initial  abundance of  helium.  This  can be  understood by  examining
Fig.~\ref{hfc}, which displays  the temporal evolution of  the mass of
the HFC from the  onset of core helium burning to  the moment at which
the first thermal pulse  is reached for the $ 1.0\,M_{\sun}$, $Z$=0.001
sequences in which  $Y$=0.4   and  $Y$=0.247 are
considered. The color scale on the  right of the figure shows the core
helium abundance. Despite the HFC mass before the onset of core helium
burning is  smaller for  the helium-enhanced  sequence, it  grows more
rapidly during  the helium burning phase  than in the case  in which a
standard  initial  helium abundance  is  considered.  This is  because
hydrogen  burning  by   the  CNO  cycle  is  more   efficient  in  the
helium-enhanced sequence.  Thus, after helium exhaustion  in the core,
the helium-enhanced sequence ends up with a larger HFC.

We now  compare the predictions  for the  final mass of  our sequences
with  those  of  \cite{2015A&A...578A.117C}.  In  particular,  we  pay
special attention to the $0.8 \,M_{\sun}$, $Z$=0.0005 sequence studied
in detail by  these authors.  As can be seen  in Fig.\ref{mimfull} for
our $ 0.8 \,M_{\sun}$, $Z$=0.0005 progenitor we obtain a mass of $\sim
0.56  \,M_{\sun}$, which  is  markedly smaller  than  that derived  by
\cite{2015A&A...578A.117C}  for  the  same initial  helium  abundance,
$0.659 \,M_{\sun}$.  This  discrepancy is mostly due  to the different
mass-loss rates used in both  studies, particularly during the AGB and
early TP-AGB  phases. In fact,  as mentioned in  Sect.~\ref{code}, for
those   phases    we   rely   on   the    mass-loss   formulation   of
\cite{2005ApJ...630L..73S} --  which is  the same we  use for  the RGB
phase.  In contrast, \cite{2015A&A...578A.117C} employed the mass-loss
rate  of  \cite{1993ApJ...413..641V}  after  the  end  of  core-helium
burning. This treatment predicts mass-loss rates substantially smaller
than those of \cite{2005ApJ...630L..73S}  during these phases. This is
particularly true for low-mass stars.  This results in a larger number
of thermal  pulses and a  larger final  mass when compared  with those
obtained using our  prescriptions. To check this,  we have re-computed
our  $  0.85  \,M_{\sun}$,  $Z$=0.0005 sequence,  but  now  using  the
mass-loss rate  of \cite{1993ApJ...413..641V} after the  end of helium
core burning.   During the AGB  and early TP-AGB, the  mass-loss rates
are between  2 and 3 orders  of magnitude smaller than  those given by
\cite{2005ApJ...630L..73S}.   As a  result,  our sequence  experiences
many more thermal pulses (20) and ends its evolution with a final mass
of $0.675  \,M_{\sun}$, markedly larger  than the final mass  of $0.58
\,M_{\sun}$     we      find     when     the      prescription     of
\cite{2005ApJ...630L..73S}  is used.   This explains  the much  larger
final mass found in  \cite{2015A&A...578A.117C}, as compared with that
obtained  here.  However,  we  mention  that for  the  AGB phase,  the
prescription  of \cite{1993ApJ...413..641V}  yields  a mass-loss  rate
much smaller  than expected  for the RGB  phase. This  contradicts the
mass-loss   estimates   obtained   for  stars   in   $\omega$~Centauri
\citep{2009MNRAS.394..831M}.  For  instance, at $\log(L/L_{\sun})=3.0$
on  the AGB  the prescription  of \cite{1993ApJ...413..641V}  yields a
mass-loss rate  three orders  of magnitud times  smaller than  that of
\cite{2005ApJ...630L..73S} for the RGB  at the same luminosity.  Also,
\cite{2014ApJ...790...22R}  have shown  that  the  mass-loss rates  of
\cite{2005ApJ...630L..73S}   provide  a   consistent  description   of
pre-dust AGB winds.

The   theoretical    initial-to-final   mass   relations    shown   in
Fig.~\ref{mimfull}  correspond  to  evolutionary sequences  for  which
overshooting was disregarded during the TP-AGB phase. For all of these
sequences, we did not find  third dredge-up episodes.  However, at low
metallicities overshooting  favors the  occurrence of  third dredge-up
episodes and  carbon enrichement of  the envelope also for  very small
stellar masses \citep{2009A&A...508.1343W}.  Nevertheless, there is no
conclusive evidence  about the occurrence of  third dredge-up episodes
in metal-poor,  low-mass progenitors  \citep{2015A&A...576A...9A}.  To
assess the impact of overshooting during  the TP-AGB on the final mass
of the remnant we have re-computed  some of our sequences for $Y$=0.4,
$Z$=0.001  but  allowing  overshooting   from  the  beginning  of  the
thermally-pulsing   AGB  phase.    Results   are   depicted  in   Fig.
\ref{mimfull}     with    a     blue    dotted     line.     Following
\cite{2016A&A...588A..25M}, we considered  diffusive overshooting with
the    overshooting   parameter    set    to    $f$=0.0075   --    see
\cite{2010ApJ...717..183R, 2009A&A...508.1343W} for details.  As shown
in \cite{2016A&A...588A..25M}, and  mentioned in Sect.~\ref{code}, the
choice of  $f$=0.0075 reproduces  several observational  properties of
AGB and  post-AGB stars in  our Galaxy  and in the  Magellanic Clouds.
The  occurrence of  third  dredge-up episodes  in  the sequences  with
overshooting reduce  further growth  of the  HFC, thus  yielding lower
final masses.   Also, the increase in  the C/O ratio resulting  from a
third  dredge-up  episode results  in  cooler  stars. This,  in  turn,
translates in significantly  larger mass-loss rates. As  a result, the
remnant of the evolution departs from the AGB after a few more thermal
pulses following the occurrence of third dredge-up.  We find that only
our 1.5 and $2.0 \,M_{\sun}$ sequences experience third dredge-up.  In
particular,  for the  $2.0 \,M_{\sun}$  sequence, C/O  $>1$ after  the
third thermal  pulse, and becomes  C/O $>2.5$ after the  tenth thermal
pulse.  From  Fig.~\ref{mimfull} we conlcude that  the intial-to-final
mass relation  for helium-enhanced stars  is not markedly  modified by
the occurrence of overshooting during the TP-AGB phase.

\begin{figure}
\centering
\includegraphics[clip,width=\columnwidth]{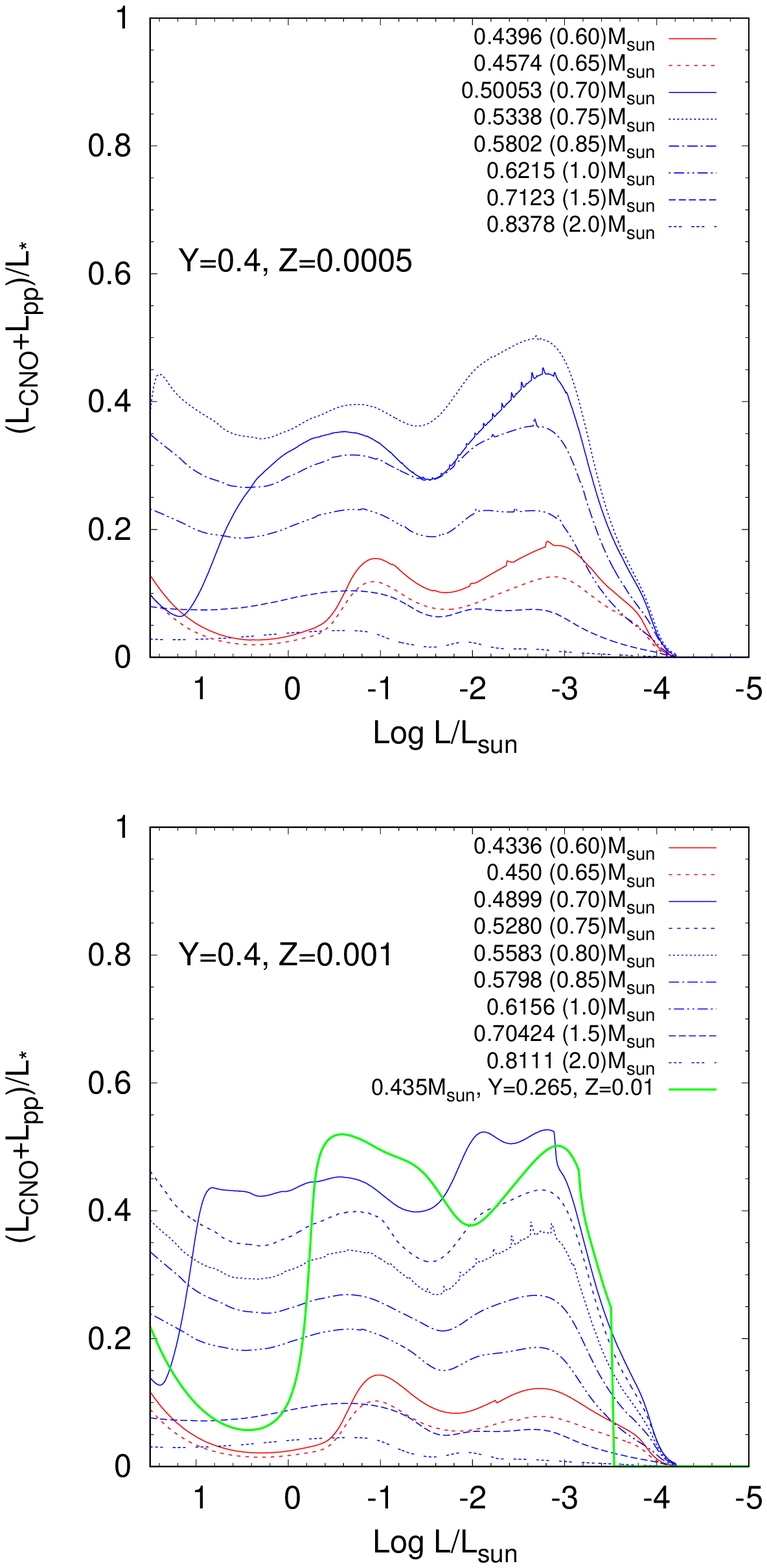}
\caption{Fraction of the total white  dwarf luminosity due to hydrogen
  nuclear burning (CNO  and $pp$ burning) for  all our helium-enhanced
  sequences  for metallicity  $Z$=0.0005  and $Z$=0.001  -- upper  and
  bottom  panels, respectively  --  and initial  helium mass  fraction
  $Y$=0.4. Red (blue) lines show the predictions of our sequences that
  end  as  white  dwarfs  with  helium (C/O)  cores.  The  green  line
  corresponds  to  a  helium-core   white  dwarf  sequence  of  $0.435
  \,M_{\sun}$ resulting  from a progenitor with  $Z$=0.01 and standard
  initial         helium         abundance,         taken         from
  \cite{2013A&A...557A..19A}.  Note that,  for  the  less massive  C/O
  white dwarfs,  residual hydrogen  burning becomes a  relevant energy
  source even at low luminosities.}
\label{lfrac}
\end{figure}

\subsection{The evolution of white dwarfs}

We now pay  attention to the evolution of white  dwarfs resulting from
helium-enhanced  progenitors.   We begin  examining  Fig.~\ref{lfrac},
where we plot the fraction of  the total white dwarf luminosity due to
hydrogen  nuclear burning  (CNO and  $pp$ burning)  during the  entire
cooling phase  for all  our helium-enhanced evolutionary  sequences of
metallicity  $Z$=0.0005  and  $Z$=0.001,   upper  and  bottom  panels,
respectively,  and  initial helium  mass  fraction  $Y$=0.4.  In  this
figure, the  red lines show  the predictions for those  sequences that
end as helium-core white dwarfs,  whilst blue lines depict the results
for those  sequences which result in  white dwarfs with C/O  cores. In
addition, in the bottom panel of this figure we include the prediction
for  a helium-core  white dwarf  sequence of  mass $0.435  \,M_{\sun}$
resulting  from  the  evolution  of a  progenitor  with  $Z$=0.01  and
standard initial  helium abundance  \citep{2013A&A...557A..19A}.  Note
that, for the less massive C/O white dwarfs, residual hydrogen burning
becomes a  main energy source  even at  low luminosities.  This  is in
line    with   the    results   of    \cite{2013ApJ...775L..22M}   and
\cite{2015A&A...576A...9A},  who  found  that  low-mass  white  dwarfs
resulting from  low-metallicity progenitors that have  not experienced
third dredge-up, are charaterized by massive hydrogen envelopes. This,
in turn, has one evident effect, namely that residual hydrogen burning
becomes one of  the main energy sources during the  evolution of white
dwarfs for  substantial periods  of time.  The  calculations performed
here  show that  this is  also true  for white  dwarfs resulting  from
helium-enhanced progenitors.

As     shown      in     \cite{2015A&A...576A...9A},      see     also
\cite{2016A&A...588A..25M},  for low-mass,  low-metallicity progenitor
stars overshooting during the TP-AGB  phase leads to carbon enrichment
of  the envelope  due to  third dredge-up  episodes, and  consequently
reduces  the  final mass  of  the  hydrogen  envelope with  which  the
corresponding  white  dwarfs enter  into  their  cooling track.  This,
obviously,  minimizes the  role played  by residual  hydrogen burning.
Hence,   the  conclusion   reached  by\cite{2015A&A...576A...9A}   and
\cite{2016A&A...588A..25M} that residual  hydrogen burning impacts the
cooling of low-metallicity white dwarfs is not valid if some amount of
overshooting  is   allowed  during   the  TP-AGB  phase   of  low-mass
progenitors.  This is in contrast with the situation we find for white
dwarfs  resulting  from  helium-enhanced   progenitors.  In  fact,  as
mentioned  previously,  when  overshooting is  considered  during  the
TP-AGB  phase  of  helium-enhanced progenitors  third  dredge-up  only
occurs for our more massive sequences.   For the less massive ones, no
carbon enrichement  is predicted when overshooting  is considered, and
consequently residual hydrogen  burning plays a role  in the evolution
of  the resulting  white  dwarfs. To  summarize,  the conclusion  that
stable hydrogen burning dominates a  significant part of the evolution
of low-mass  white dwarfs  resulting from  low-metallicity He-enhanced
progenitors does not  depend on the occurrence  of overshooting during
the TP-AGB evolution.

Another interesting  feature of the results  shown in Fig.~\ref{lfrac}
is the small  impact of residual hydrogen burning on  the evolution of
massive  helium-core  white   dwarfs  resulting  from  helium-enhanced
progenitors. Again,  this is because  these white dwarfs  experience a
CNO shell flash during the cooling branch that reduces the mass of the
residual hydrogen-rich envelope, thus  minimizing the role of hydrogen
burning at advanced stages of the  evolution. This is in contrast with
what occurs  for helium-core  white dwarfs resulting  from progenitors
with     standard     initial     helium     abundances     --     see
\cite{2014A&A...571L...3I}  and \cite{2013A&A...557A..19A}  for recent
works -- since  for very massive helium-core white  dwarfs CNO flashes
do not  occur, even when element  diffusion is taken into  account. To
illustrate this in Fig.~\ref{lfrac} we show, using a thick green line,
the contribution of residual hydrogen  burning for a helium-core white
dwarf of mass $0.435 \,M_{\sun}$ resulting from a progenitor star with
$Z$=0.01     and    a     standard     initial    helium     abundance
\citep{2013A&A...557A..19A}.   For  this  star, no  CNO  flashes  took
place,   and  thus   in  this   case,  hydrogen   burning  contributes
significantly to the luminosity of the white dwarf.

\begin{figure}
\centering
\includegraphics[clip,width=\columnwidth]{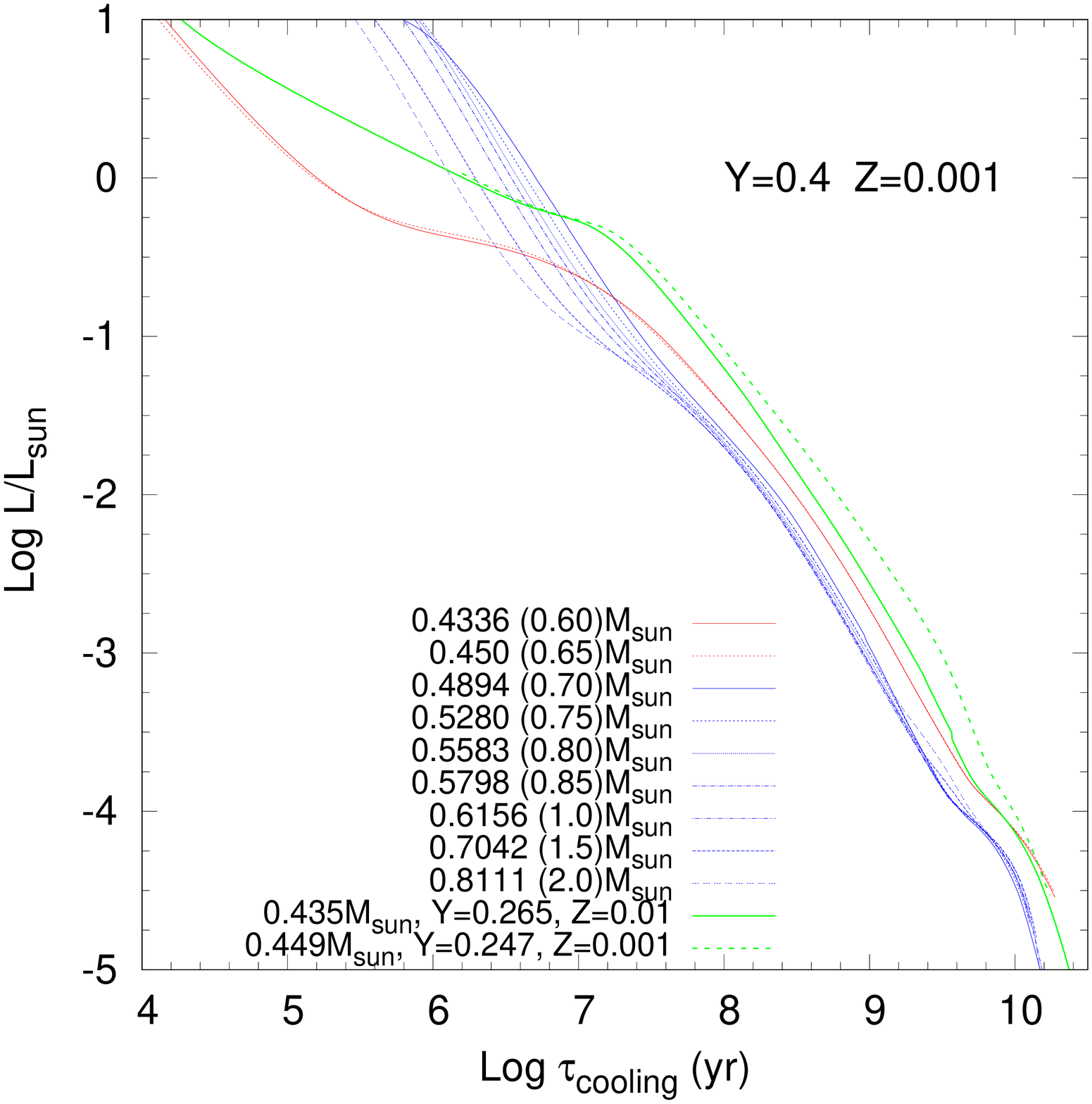}
\caption{Cooling  times for  all our  helium-enhanced sequencies  with
  $Z$=0.001 and  initial helium mass  fractions $Y$=0.4. Red  and blue
  lines correspond, respectively, to white  dwarfs with helium and C/O
  cores.   The solid  green line  corresponds to  a helium-core  white
  dwarf of  mass $0.435 \,M_{\sun}$  resulting from a  progenitor with
  $Z$=0.01     and      standard     initial      helium     abundance
  \citep{2013A&A...557A..19A},   whilst   the    dashed   green   line
  corresponds to  helium-core white dwarfs of  mass $0.449 \,M_{\sun}$
  resulting  from a  progenitor  with $Z$=0.001  and standard  initial
  helium abundance \citep{2002MNRAS.337.1091S}. }
\label{edad_Y04Z1d3}
\end{figure}

\begin{figure}
\centering
\includegraphics[clip,width=\columnwidth]{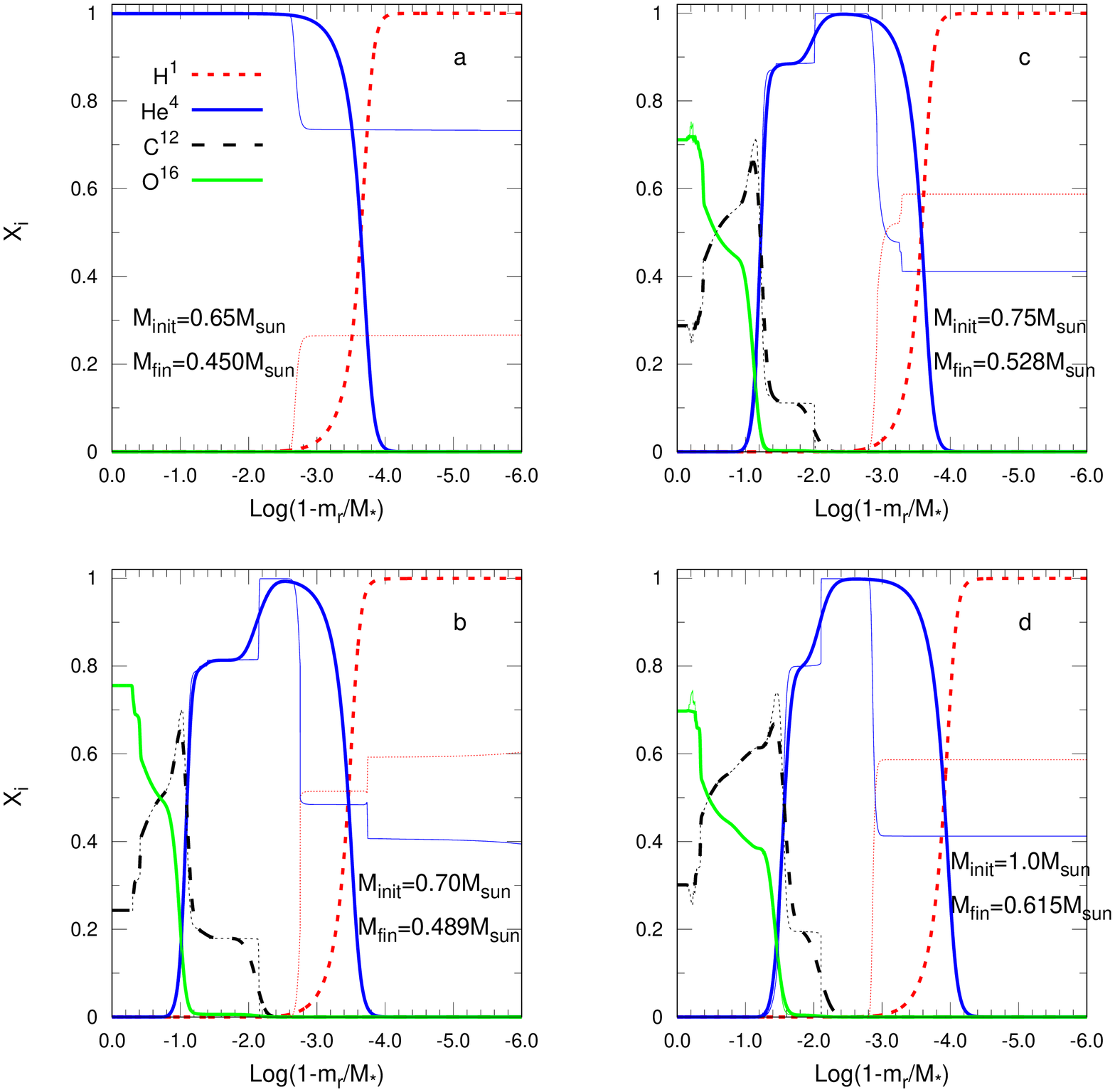}
\caption{Chemical abundance  distribution of hydrogen,  helium, carbon
  and oxygen  in terms of the  outer mass fraction for  selected white
  dwarf  models at  the  ZZ~Ceti stage  (thick  lines) resulting  from
  helium-enhanced  progenitors  with   $Z$=0.001  and  initial  helium
  abundance $Y$=0.4.  The chemical  profiles at  the beginning  of the
  cooling  track  are  the  thin  lines.  Panel  a  corresponds  to  a
  helium-core  white dwarf  resulting from  a progenitor  of initially
  $M=0.65\,M_{\sun}$, while  panels b,  c, and d  to C/O  white dwarfs
  resulting     from      progenitors     with     $M=0.70\,M_{\sun}$,
  $M=0.75\,M_{\sun}$, and $M=1.0\,M_{\sun}$, respectively.}
\label{perfiles}
\end{figure}

The resulting cooling times  are displayed in Fig.~\ref{edad_Y04Z1d3}.
This  figure   shows  the   cooling  times   of  our   sequencies  for
helium-enhanced progenitors  with $Z$=0.001, $Y$=0.4. The  time origin
is  taken at  the moment  at  which the  remnants reach  the point  of
maximum effective temperature at the  beginning of the cooling branch.
As in Fig.~\ref{lfrac}, we show with blue lines, the results for white
dwarfs with C/O cores, and with red lines, those for white dwarfs with
helium cores. We find that residual nuclear burning yields substantial
delays in the cooling time for  the less massive C/O white dwarfs.  At
$\log(L/L_{\sun}) \sim  -3.0$, residual  hydrogen burning leads  to an
increase in  the cooling times  of low-mass white dwarfs  which ranges
between 20  and 40\%.  This  conclusion is independent of  the assumed
initial helium abundance.  In fact, \cite{2015A&A...576A...9A} found a
similar result for the case of white dwarfs resulting from progenitors
with standard  helium abundances.   However, as mentioned  earlier, in
contrast with  what is found for  the case of standard  sequences, the
impact of residual  hydrogen burning on the cooling  times of low-mass
white  dwarfs  resulting  from   helium-enhanced  progenitors  is  not
affected  by the  occurrence of  overshooting on  the TP-AGB  phase of
their progenitor stars.

The initial helium abundance also impacts the evolution of helium-core
white dwarfs. To show this,  in Fig.~\ref{edad_Y04Z1d3} we include the
cooling  curves  for   a  helium-core  white  dwarf   of  mass  $0.435
\,M_{\sun}$  resulting from  a progenitor  with $Z$=0.01  and standard
initial  helium  abundance   \citep{2013A&A...557A..19A},  and  for  a
helium-core white  dwarf of mass  $0.449 \,M_{\sun}$ resulting  from a
progenitor  with $Z$=0.001  and  a standard  initial helium  abundance
\citep{2002MNRAS.337.1091S},   solid    and   dashed    green   lines,
respectively.  Note  that the  helium-core white dwarf  resulting from
helium-enhanced   progenitors   evolves   much   faster   than   their
counterparts resulting  from standard progenitors. As  mentioned, this
different behavior  is due to the  occurrence of a CNO  shell flash in
the  helium-core white  dwarf  with  helium-enhanced progenitors  that
reduces the impact of further  nuclear burning at advanced stages.  In
fact,  at intermediate  and  low luminosities,  the  cooling times  of
massive helium-core white dwarfs  with helium-enhanced progenitors are
about a factor of 2 shorter,  when compared with the helium-core white
dwarfs   formed  from   progenitors  with   standard  initial   helium
abundances.

\subsection{The pulsational properties}

As already  shown, the  helium enhancement of  progenitor stars  has a
significant impact  on the  formation and  evolution of  the resulting
white dwarfs.   In this sense,  it is  worthwhile to also  explore the
pulsational  properties  of  these   white  dwarfs.   The  pulsational
properties of these  stars are strongly dependent on the  shape of the
chemical profile left by the  evolution of their progenitors.  In this
connection,  in Fig.~\ref{perfiles}  we show,  using thick  lines, the
inner abundance distribution of hydrogen, helium, carbon and oxygen as
a function of the outer mass  fraction for selected white dwarf models
resulting from helium-enhanced progenitors  with $Z$=0.001 and initial
helium abundance $Y$=0.4 at the  ZZ~Ceti stage.  In addition, the thin
lines  show the  chemical profiles  at  the beginning  of the  cooling
track.  Panel a corresponds to  a helium-core white dwarf that results
from a progenitor of initial  mass $M=0.65\,M_{\sun}$, while panels b,
c, and  d to  C/O white  dwarfs resulting  from progenitors  of masses
$M=0.70\,M_{\sun}$,    $M=0.75\,M_{\sun}$,   and    $M=1.0\,M_{\sun}$,
respectively.   The  carbon and  oxygen  profiles  are the  result  of
different processes acting during the  evolution of the progenitor. In
particular, note the flat profile in the inner core left by convection
during  the  core helium  burning  phase  and  the signatures  of  the
outward-moving helium burning shell, particularly for the more massive
sequence  illustrated,   which  reached  the  thermally   pulsing  AGB
phase.  In addition,  the chemical  profiles at  the beginning  of the
cooling track of the C/O sequences show the typical intershell rich in
helium and carbon left by  the pulse-driven convection zone during the
thermally pulsing  phase.  This intershell  is present in all  our C/O
white dwarf sequences.  We remind that although the $M=0.70\,M_{\sun}$
and  $M=0.75\,M_{\sun}$ progenitors  avoided  the  TP-AGB phase,  they
nonetheless  experienced  several  thermal pulses  at  high  effective
temperatures before  entering the  white dwarf  stage.  The  action of
element  diffusion  is  clearly  also noticeable,  and  leads  to  the
formation  of a  thick hydrogen  envelope  by the  time evolution  has
reached  the domain  of the  ZZ~Ceti stars.  But more  importantly, it
removes the double-layered structure at the interhell region for white
dwarfs with  masses larger  than $M  \simeq 0.60\,M_{\sun}$.  For less
massive  white  dwarfs,  the  intershell  region  is  not  removed  by
diffusion.  In this  regards,  it  is important  to  realize that  the
presence  of  a  double-layered   structure  affects  the  theoretical
$g-$mode    period     spectrum    of     ZZ    Ceti     stars,    see
\cite{2010ApJ...717..897A}.

\begin{figure}
\centering
\includegraphics[clip,width=\columnwidth]{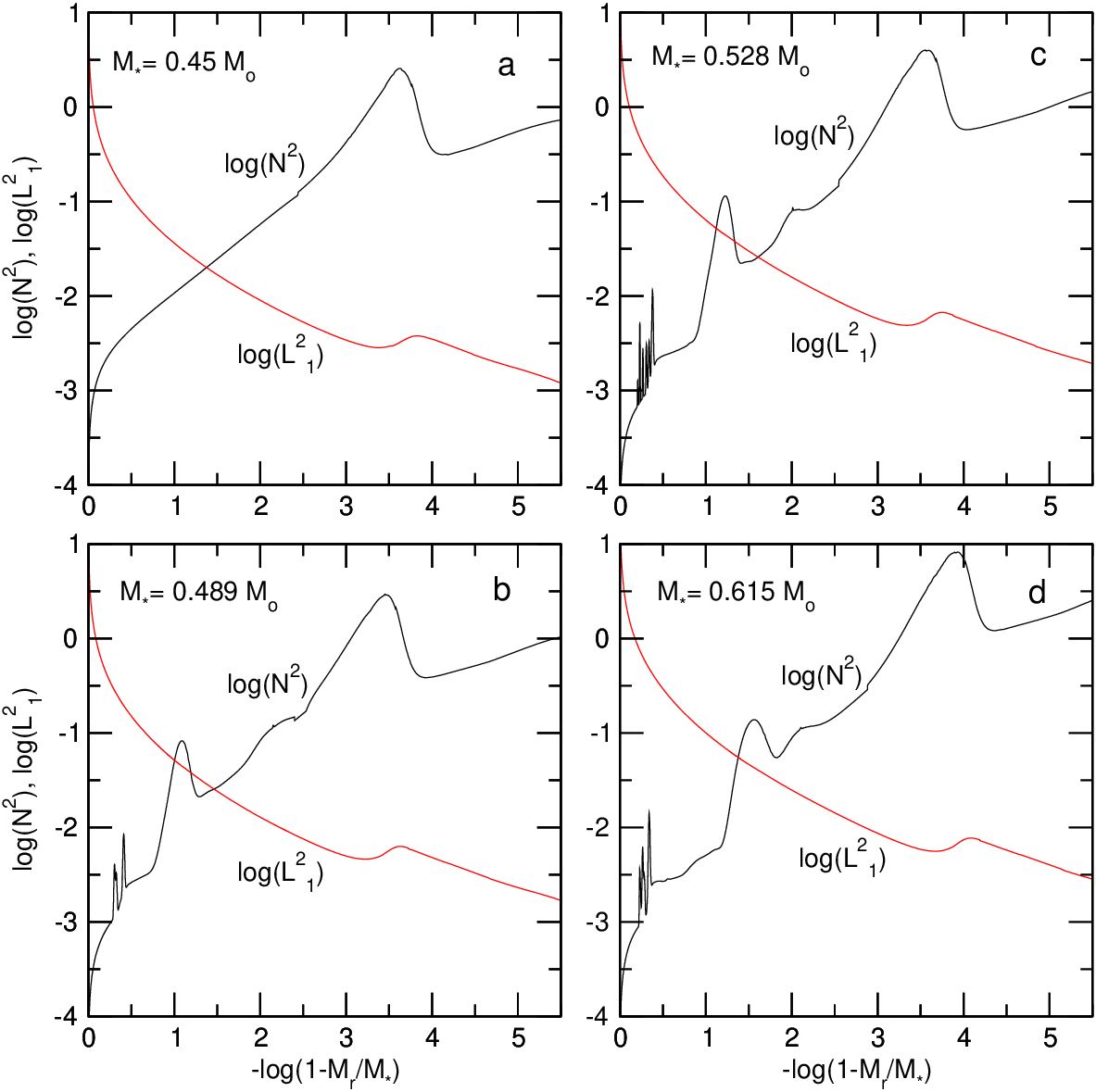}
\caption{Propagation diagrams --  the spatial run of  the logarithm of
  the  squared  Brunt-V\"ais\"al\"a  and  Lamb  frequencies  (Unno  et
  al.  1989) --  corresponding to  the  white dwarfs  models shown  in
  Fig.~\ref{perfiles} for $\ell= 1$.}
\label{bv}
\end{figure}

Once we have studied the  chemical stratification of these white dwarf
models  we continue  our  analysis computing  the adiabatic  pulsation
periods of  nonradial $g$ (gravity)  modes employing the  {\tt LP-PUL}
pulsation   code   described   in   \citet{2006A&A...454..863C}.    In
Fig.~\ref{bv}      we      show     the      propagation      diagrams
\citep{1989nos..book.....U} corresponding  to the same  stellar models
depicted  in Fig.~\ref{perfiles}.  The  propagation  diagrams are  the
spatial run of the  Brunt-V\"ais\"al\"a (buoyancy) frequency, $N$, and
the  Lamb   (acoustic)  frequency,  $L_{\ell}$.   The   shape  of  the
Brunt-V\"ais\"al\"a frequency largely determines the properties of the
$g$-mode                       pulsation                      spectrum
\citep{2008ARA&A..46..157W, 2008PASP..120.1043F, 2010A&ARv..18..471A}.
The profile of $N$ reflects all changes in chemical composition of the
white  dwarf  model in  the  form  of  local  maxima of  the  buoyancy
frequency \citep{1989nos..book.....U}. In the  case of the helium-core
white dwarf  model, there  exists only one  chemical interface  -- the
helium-hydrogen   transition    region   (see   panel   a    of   Fig.
\ref{perfiles}). This  chemical interface induces  a bump in  $N^2$ at
$-\log(1-M_r/M_{\star}) \sim  3.5$ (panel  a of Fig.   \ref{bv}).  Note
that this interface also affects the  shape of the Lamb frequency.  In
the  case  of white  dwarfs  with  carbon-oxygen cores,  the  internal
chemical structure is  substantially more complex than in  the case of
the helium-core  white dwarf models  (see panels b,  c, and d  in Fig.
\ref{perfiles}). This results  in a much more complex form  of the run
$N^2$, as can be seen in panels b, c, and d of Fig.~\ref{bv}.  Indeed,
in these cases, the  Brunt-V\"ais\"al\"a frequency is characterized by
additional maxima  induced by the presence  of the C/O-He and  the C/O
chemical interfaces.

\begin{figure}
\centering
\includegraphics[clip,width=\columnwidth]{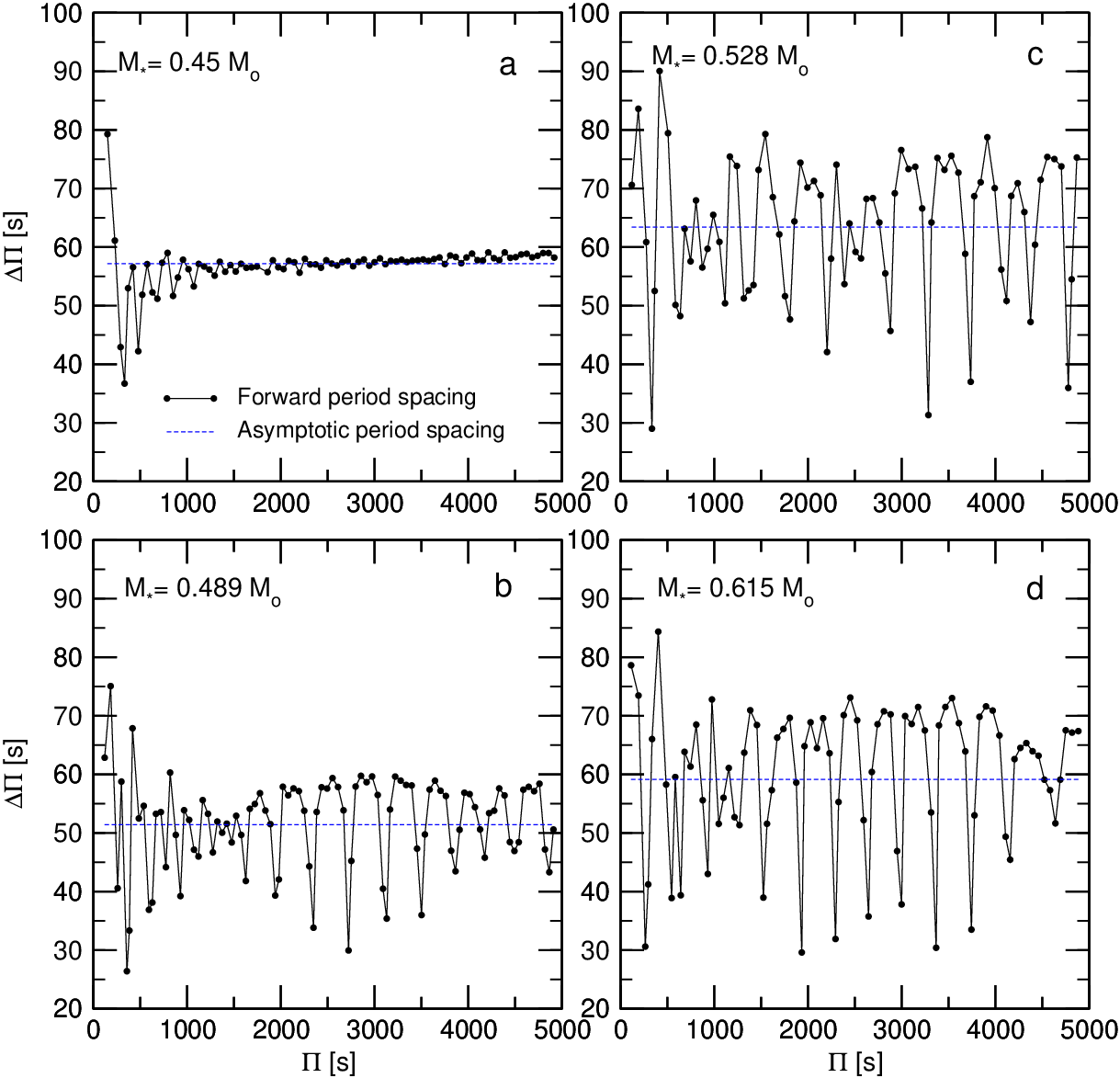}
\caption{The dipole ($\ell= 1$) forward period spacing ($\Delta Pi_k$)
  as a function  of the periods ($\Pi_k$) corresponding  to the models
  shown in Fig.~\ref{perfiles}.}
\label{dp}
\end{figure}

The  number  and shape  of  the  chemical  interfaces present  in  the
interior  of DA  white dwarf  models strongly  affect the  propagation
properties  of nonradial  pulsation $g$  modes, in  particular through
mode               trapping              and               confinement
\citep{1992ApJS...80..369B,B96,2002A&A...387..531C}.  Mode-trapping or
confinement  results  in  strong  departures from  uniformity  of  the
forward period separation, $\Delta \Pi_k$ ($\equiv \Pi_{k+1}- \Pi_k$),
when plotted in  terms of the pulsation period $\Pi_k$  ($k$ being the
radial  order of  the mode).  Thus, the  period difference  between an
observed mode and adjacent modes can be considered as an observational
diagnostic  of  mode  trapping.  For  a  helium-core  white  dwarf  --
characterized by a single chemical interface -- like that of the model
of mass  $M=0.45 M_{\sun}$  we are considering  here, local  minima in
$\Delta \Pi_k$  usually correspond  to modes  trapped in  the hydrogen
envelope, in  contrast to  local maxima in  $\Delta \Pi_k$,  which are
associated with modes  trapped in the inner core.   The forward period
spacing for $\ell= 1$ $g$ modes  in terms of the periods corresponding
to  the helium-core  model of  mass $M=0.45  M_{\sun}$ is  depicted in
panel  a  of Fig.~\ref{dp}  .   We  also  show the  asymptotic  period
spacing, computed as in  \cite{1990ApJS...72..335T} with a dotted blue
line.  Mode-trapping  signatures are clearly  noticeable, particularly
for periods  shorter than $\sim  1500$~s.  Longer periods seem  to fit
the asymptotic  predictions, although  small departures  from constant
period spacing  are still clearly seen.   In the case of  white dwarfs
with carbon-oxygen cores, the presence of multiple chemical transition
regions     causes      much     more     complex      patterns     of
mode-trapping/confinement, as  can be seen  in panels  b, c, and  d of
Fig.~\ref{dp}. In  particular, strong  departures from  uniform period
separation  are  evident for  the  models  with $M  =0.489\,M_{\sun}$,
$M=0.528\, M_{\sun}$, and $M=0.615\, M_{\sun}$. These features persist
for  the entire  range of  periods analyzed  here. We  also note  some
``beating''  modulating the  amplitudes of  the departures  of $\Delta
\Pi_k$. This beating is  due to the combined mode-trapping/confinement
effects caused by the various steps in  the C/O profile in the core --
see  \cite{2006A&A...454..863C} for  the case  of PG1159  star models.
Clearly,  the mode  trapping/confinement features  of models  with C/O
cores are by far more pronounced  than for the case of the helium-core
white dwarf.

\section{Summary and conclusions}
\label{conclusions}

In this  work we have  computed the full evolution  of helium-enhanced
($Y$=0.4) sequences  with initial masses  ranging from 0.60  to $2.0\,
M_{\sun}$.  Two  sets of model  sequences have been computed,  for the
first one we  adopted a metallicity $Z$=0.001, whereas  for the second
one  $Z$=0.0005  was employed.   Emphasis  has  been placed  on  those
aspects  of  evolution  of   helium-enhanced  star  relevant  for  the
formation  and evolution  of  white dwarfs.   For  all our  sequences,
evolution  has  been computed  starting  from  the ZAMS  and  followed
through the  stages of  stable hydrogen and  helium core  burning, the
stage of mass loss during the entire thermally-pulsing AGB, the domain
of the  planetary nebulae at  high effective temperature,  and finally
the  terminal  white dwarf  cooling  track,  up  to very  low  surface
luminosities.  To the  best of our knowledge,  the sequences presented
here constitute the  first set of consistent  evolutionary tracks from
the  ZAMS  to  the  white  dwarf   stage  covering  a  wide  range  of
masses. This  set of  sequences is  appropriate for  the study  of the
formation and evolution of white dwarfs resulting from  stars with
very high initial helium abundances,  like  those   found  in $\omega$~Centauri
and NGC~2808.

We explored different aspects of  the evolution of the resulting white
dwarfs  and  of  their  progenitors. In  particular,  we  studied  the
initial-to-final mass  relation, and we  found that the final  mass of
white dwarfs  resulting from  helium-enhanced progenitors  is markedly
larger than  the final  mass expected from  sequences with  a standard
helium initial abundance.  Progenitors  with initial mass smaller than
$M  \simeq  0.65\,M_{\sun}$ evolve  directly  into  white dwarfs  with
helium cores  in less  than 14~Gyr, while  those descending  from more
massive  progenitors  and  with  initial  stellar  mass  smaller  than
$M\simeq  0.75\,M_{\sun}$, reach  the white  dwarf stage  avoiding the
TP-AGB phase, TP-AGB manqu\`e (or AGB  manqu\`e if the stellar mass is
less than $M$$\simeq$$ 0.70\,M_{\sun}$).  However, before reaching the
terminal  cooling  track,  these remnants  experience  several  helium
thermal pulses  at high  effective temperatures.   It is  worth noting
that the formation  of helium-core white dwarfs in  our simulations is
not the  result of  binary evolution,  but a  consequence of  the much
shorter ages  of the  low-mass, helium-enhanced progenitors.   This is in
accordance   with  the   results  of   \cite{2008ApJ...673L..29C}, 
\cite{2009ApJ...702.1530C}, and \cite{2013ApJ...769L..32B}, who found 
evidence for the existence of an important fraction of helium-core 
white dwarfs in $\omega$~Centauri.

Our  simulations show  that for  initial  masses larger  than $M  \sim
1.0\,M_{\sun}$  the hydrogen-free  core grows  during the  TP-AGB. For
lower   stellar  masses,   the   final  mass   of   the  remnants   of
helium-enhanced progenitors does not  differ appreciably from the mass
of the  hydrogen-free core at  the first  thermal pulse. We  have also
explored the impact of overshooting  during the TP-AGB, and found that
the occurrence of third dredge-up and the related carbon enrichment of
the envelope is expected to occur for initial masses larger than $\sim
1.5  \,M_{\sun}$.  However,  we  found that  the intial-to-final  mass
relation  for helium-enhanced  stars is  not markedly  altered by  the
occurrence of overshooting during the TP-AGB phase.

With regard  to the evolution of  the resulting white dwarfs,  we find
that for the less massive  C/O white dwarfs, residual hydrogen burning
becomes a  main energy source  even at low luminosities,  in agreement
with  the predictions  of \cite{2015A&A...576A...9A},  who found  that
low-mass white dwarfs resulting  from low-metallicity progenitors that
have  not experienced  third dredge-up,  are charaterized  by residual
hydrogen burning,  so this energy  source dominates the  evolution for
substantial periods of time. The resulting delays in the cooling times
range from  20 to  $40\%$ at low  luminosities.  However,  in contrast
with  the  situation  encountered  for  white  dwarfs  resulting  from
standard progenitors, we found that the role played by stable hydrogen
burning  in  low-mass  white dwarfs  descending  from  helium-enhanced
progenitors does not  depend on the occurrence  of overshooting during
the TP-AGB evolution.  In addition,  according to our simulations, the
very massive helium-core white  dwarfs descending from helium-enhanced
progenitors experience  a CNO shell  flash during the  cooling branch,
thus minimizing the  impact of hydrogen burning at  advanced stages of
the   evolution.   As   a   consequence,  at   intermediate  and   low
luminosities, cooling times of massive helium-core white dwarfs formed
from helium-enhanced  progenitors are  about a factor  of 2  lower, as
compared with helium-core white dwarfs resulting from progenitors with
a standard initial helium abundance.

We also  find that  a double-layered structure  in the  outer chemical
profile   is  formed   in  all   the  resulting   C/O  models.    This
double-layered chemical  profile is  the result of  the mixing  in the
pulse-driven convection  zone that appears  as a result of  the helium
thermal pulses.  However, by the  time the ZZ~Ceti regime  is reached,
this structure is removed by  element diffusion for masses larger than
$M  \simeq 0.60\,M_{\sun}$.   This  behavior  impacts the  pulsational
properties of the models at the ZZ~Ceti stage.

Finally, the pulsational properties of  our models have been explored.
We have  found that  in the  case of  white dwarfs  with carbon-oxygen
cores, the shape of  the Brunt-V\"ais\"al\"a frequency is substantially
more complex than in the case  of white dwarfs with helium cores. This
leads     to    a     markedly    different     behavior    of     the
mode-trapping/confinement   properties   and    the   forward   period
separation, $\Delta  \Pi_k$ ($\equiv  \Pi_{k+1}- \Pi_k$). In  spite of
the  similarity  of  the  stellar   masses  analyzed  here,  the  mode
trapping/confinement   features  of   models   with   C/O  cores   are
substantially different from those models with helium cores.  Then, if
pulsations  are detected  in future  photometric observations  in this
type of  objects, this  distinctive difference  in the  $\Delta \Pi_k$
distribution  could  be  employed  as a  seismic  diagnostic  tool  to
distinguish white dwarfs with helium  cores from those with C/O cores,
provided that enough  consecutive pulsation periods of  $g$ modes were
detected.   This  constitutes  a  promising avenue  to  constrain  the
evolutionary history of cluster stars.

\begin{acknowledgements}
We thank our referee for his/her valuable comments. Part of  this work 
was  supported by  AGENCIA through the  Programa de
Modernizaci\'on    Tecnol\'ogica   BID    1728/OC-AR,   by    the   PIP
112-200801-00940 grant from CONICET,  by MINECO grant AYA2014-59084-P,
and by  the AGAUR.  This  research has  made use of  NASA Astrophysics
Data System.
\end{acknowledgements}

\bibliographystyle{aa}
\bibliography{He40}

\end{document}